\documentclass[useAMS,usenatbib]{mn2e}

\usepackage[active]{srcltx}
\usepackage{graphicx}
\usepackage{txfonts}
\usepackage{multirow}
\usepackage{longtable}
\usepackage[francais]{babel}
\usepackage[applemac]{inputenc}

\newcommand{\kms}{km.s$^{-1}$}

\newcommand{\vsini}{$v\sin i$}
\newcommand{\wsini}{$W\sin i$}
\newcommand{\vsinis}{$v\sin i\;$}

\newcommand{\angs}{\AA$\;$}
\newcommand{\msun}{M$_{\odot}$}

\title[A spectropolarimetric survey of HAeBe stars - II. Rotation]{A high-resolution spectropolarimetric survey of Herbig Ae/Be stars\thanks{Based on observations obtained at the Canada-France-Hawaii Telescope (CFHT) which is operated by the National Research Council of Canada, the Institut National des Sciences de l'Univers of the Centre National de la Recherche Scientifique of France, and the University of Hawaii}\\II. Rotation}
\author[E. Alecian et al.]
{E.~Alecian$^{1,2}$\thanks{E-mail: evelyne.alecian@obspm.fr},
 G.A.~Wade$^2$,
 C.~Catala$^1$,
 J.H.~Grunhut $^{2,3}$,
 J.D.~Landstreet$^{4,5}$,
 T.~B\"ohm$^{6,7}$,
 \newauthor
 C.P.~Folsom$^5$,
 S.~Marsden$^{8,9}$ \\
 $^1$LESIA-Observatoire de Paris, CNRS, UPMC Univ., Univ. Paris-Diderot, 5 place Jules Janssen, F-92195 Meudon Principal Cedex, France, \\
 $^2$Dept. of Physics, Royal Military College of Canada, PO Box 17000, Stn Forces, Kingston K7K 7B4, Canada \\
 $^3$Department of Physics, Queen's University, Kingston, Canada \\
 $^4$Dept. of Physics \& Astronomy, University of Western Ontario, London N6A 3K7, Canada \\
 $^5$Armagh Observatory, College Hill, Armagh BT61 9DG, Northern Ireland, UK\\
 $^6$ Universit\'e de Toulouse; UPS-OMP; IRAP; Toulouse, France \\
 $^7$CNRS; IRAP; 14, avenue Edouard Belin, F-31400 Toulouse, France \\
  $^8$Centre for Astronomy, School of Engineering and Physical Sciences, James Cook University, Townsville, 4811, Australia \\
 $^{9}$Faculty of Sciences, University of Southern Queensland, Toowoomba, 4350, Australia
}

\begin{document}

\date{Accepted . Received ; in original form }

\pagerange{\pageref{firstpage}--\pageref{lastpage}} \pubyear{2002}

\maketitle

\label{firstpage}

\begin{abstract}
We report the analysis of the rotational properties of our sample of Herbig Ae/Be (HAeBe) and related stars for which we have obtained high-resolution spectropolarimetric observations. Using the projected rotational velocities measured at the surface of the stars, we have calculated the angular momentum of the sample and plotted it as a function of age. We have then compared the angular momentum and the \vsinis distributions of the magnetic to the non-magnetic HAeBe stars. Finally we have predicted the \vsinis of the non-magnetic, non-binary ("normal") stars in our sample when they reach the ZAMS, and compared them to various catalogues of the \vsinis of main-sequence stars. First, we observe that magnetic HAeBe stars are much slower rotators than normal stars, indicating that they have been more efficiently braked than the normal stars. In fact, the magnetic stars have already lost most of their angular momentum, despite their young ages (lower than 1 Myr for some of them). Secondly, { our analysis suggests} that the low mass ($1.5 < M < 5$ M$_{\odot}$) normal HAeBe stars evolve with constant angular momentum towards the ZAMS, while the high-mass normal HAeBe stars ($M > 5$ M$_{\odot}$) are losing angular momentum. We propose that winds, which are expected to be stronger in massive stars, are at the origin of this phenomenon.
\end{abstract}


%
%
%________________________________________________________________

\section{Introduction}

Among intermediate mass A/B stars on the main sequence (MS), a sub-group called chemically peculiar Ap/Bp stars shows very specific properties: abundance anomalies at their surface, slow rotation (rotation periods longer than 1 day), and strong magnetic fields. While the abundance anomalies are believed to result from diffusion within their surface layers due to the competition between radiative levitation and gravitational settling \citep[e.g.][]{michaud70}, the origin of the slow rotation is less well understood. The mechanism inducing the slow rotation of Ap/Bp stars is likely related to their strong magnetic fields. \citet{stepien00} discussed different theories, and he concludes that magnetic braking must occur during the pre-main sequence (PMS) phase in order to reproduce the angular momentum observations of the MS Ap/Bp stars. St{\c e}pie{\'n} demonstrated that magnetic coupling of a PMS star with its accretion disk would slow the star and reduce its rotation period to a few days. In order to produce the very slowest rotators, with observed rotation periods greater than a month, the disk must disappear sufficiently early during the PMS phase, allowing strong magnetised winds to carry away a large quantity of angular momentum until the stars reach the zero-age main-sequence (ZAMS).

The PMS progenitors of MS A/B stars are found among Herbig Ae/Be  (HAeBe) stars. These stars have emission lines in their spectra, lie in an obscured region, illuminate fairly bright nebulae, and show a pronounced infrared excess \citep{herbig60,the94,vieira03}, all of which are the result of their young age. All of them do not necessarily show all Herbig characteristics, but all of them have infrared excess with an abnormal extinction law (compared to classical Be stars). The PMS status of these stars was first confirmed by the spectroscopic study of \citet{strom72}, by showing that their surface gravities are systematically lower than those of their MS counterparts.

Until recently we had very few observational constraints on the magnetic fields and rotation of these objects. To our knowledge two thorough observational studies of the evolution of angular momentum of intermediate-mass stars during the PMS phase have been realised. \citet{bohm95}, using MUSICOS high-resolution spectra, performed a statistical study of the projected rotational velocities ($v\sin i$) of 29 HAeBe stars. They studied the evolution of the spin angular momentum of intermediate mass stars during the PMS phase by comparing the $v\sin i$ of field HAeBe stars to that of MS A/B stars in young open clusters. They concluded that, assuming solid body rotation within the star, the evolution depends on the mass of the star: low mass HAeBe stars ($M \le 2.6$~M$_{\odot}$) seem to lose a large fraction of their angular momentum during their PMS evolution; intermediate mass HAeBe stars ($2.6 \le M \le 4$~M$_{\odot}$) seem not to lose a significant fraction of their angular momentum; and higher mass HAeBe stars ($M \ge 4$~M$_{\odot}$) seem to gain a large amount of angular momentum. However, if the stellar internal rotation varies as $r^{-2}$, the observations of HAeBe stars and A/B stars in young clusters are consistent with conservation of total angular momentum at all masses.

\citet{wolff04} studied the $v\sin i$ of 145 stars between 0.4 and 14 M$_{\odot}$ located in the Orion star-forming complex. They argued that observations of the surface rotation velocities at different evolutionary stages of the PMS phase can provide information about the internal angular momentum profile of stars. They therefore distinguished between stars still on their convective track (the Hayashi phase) and more evolved stars on their radiative track. Their conclusions lead to the following scenario concerning intermediate mass stars ($M \ge 1.5$~M$_{\odot}$): stars lose angular momentum before they start the PMS phase, a decoupling of the angular momentum seen at the surface from the angular momentum in the interior occurs when the stars make the transition from convective to radiative PMS evolution, after which the angular momentum is conserved during the PMS radiative evolution. 

Both analyses gave very interesting results, and should now be discussed in the framework of a scenario of angular momentum evolution that includes magnetic fields. With this aim, we have used the large survey of 70 HAeBe stars that we performed using the high-resolution spectropolarimeters ESPaDOnS at the Canada-France-Hawaii Telescope (CFHT, Hawaii), and Narval at the Telescope Bernard Lyot (TBL, Pic du Midi, France). In a previous paper \citep[][paper I]{paperi}, we have described the whole observed sample, including the measurement of their projected rotation velocities (\vsini). We now present the second paper of the series describing the analyses of the rotational velocities and angular momentum evolution of the sample. Section 2 gives a short summary of the observations and the measurements reported in paper I. Section 3 reports our findings concerning the angular momentum evolution among our sample, while Section 4 describes the analysis performed on the projected rotational velocities. Finally, in Section 5 we conclude and discuss our results.

%
%
%________________________________________________________________

\section{Summary of the sample and observations}

\begin{figure}
\centering
\includegraphics[width=8cm]{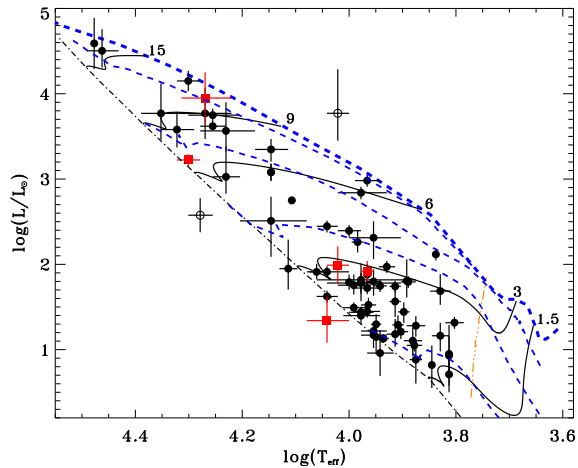}
\caption{Magnetic (red squares) and non-magnetic (black points) Herbig Ae/Be stars plotted in a Hertzsprung-Russell diagram. Open circles correspond to HD~98922 (above the birthline) and IL~Cep (below the ZAMS) that fall outside of the PMS region of the HR diagram, whose positions cannot be reproduced with the theoretical evolutionary tracks considered in this paper. The CESAM PMS evolutionary tracks for 1.5, 3, 6, 9 and 15~M$_{\odot}$ (black full lines), 0.01, 0.1, 1 and 10~Myr isochrones (blue thin dashed lines), and the ZAMS (black dot-dashed line) are also plotted. The birthline taken from \citet{behrend01} is plotted with a blue thick dashed line. The convective/radiative phase transition is overplotted with an orange dot-dot-dot-dashed line.}
\label{fig:hr}
\end{figure}

We used the ESPaDOnS and Narval instruments in polarimetric mode to obtain 132 spectra of 70 HAeBe stars with a resolving power of 65000. The data were reduced using the ‘{\sc Libre Esprit}’ package especially developed for ESPaDOnS and Narval, and installed at the CFHT and at the TBL \citep{donati97}. After reduction we obtained the intensity (Stokes $I$) and the circular polarization (Stokes $V$) spectra of the stars observed. We then applied the least-squares deconvolution (LSD) procedure, a multi-line analysis technique, allowing us to increase the signal-to-noise ratio of our data by extracting the mean Stokes $I$ and $V$ profiles for each spectrum. The Stokes $I$ profiles were then used to measure the \vsinis at the surface of each star. More details on the reduction and analysis procedures are given in paper I.

\begin{figure}
\centering
\includegraphics[width=7cm]{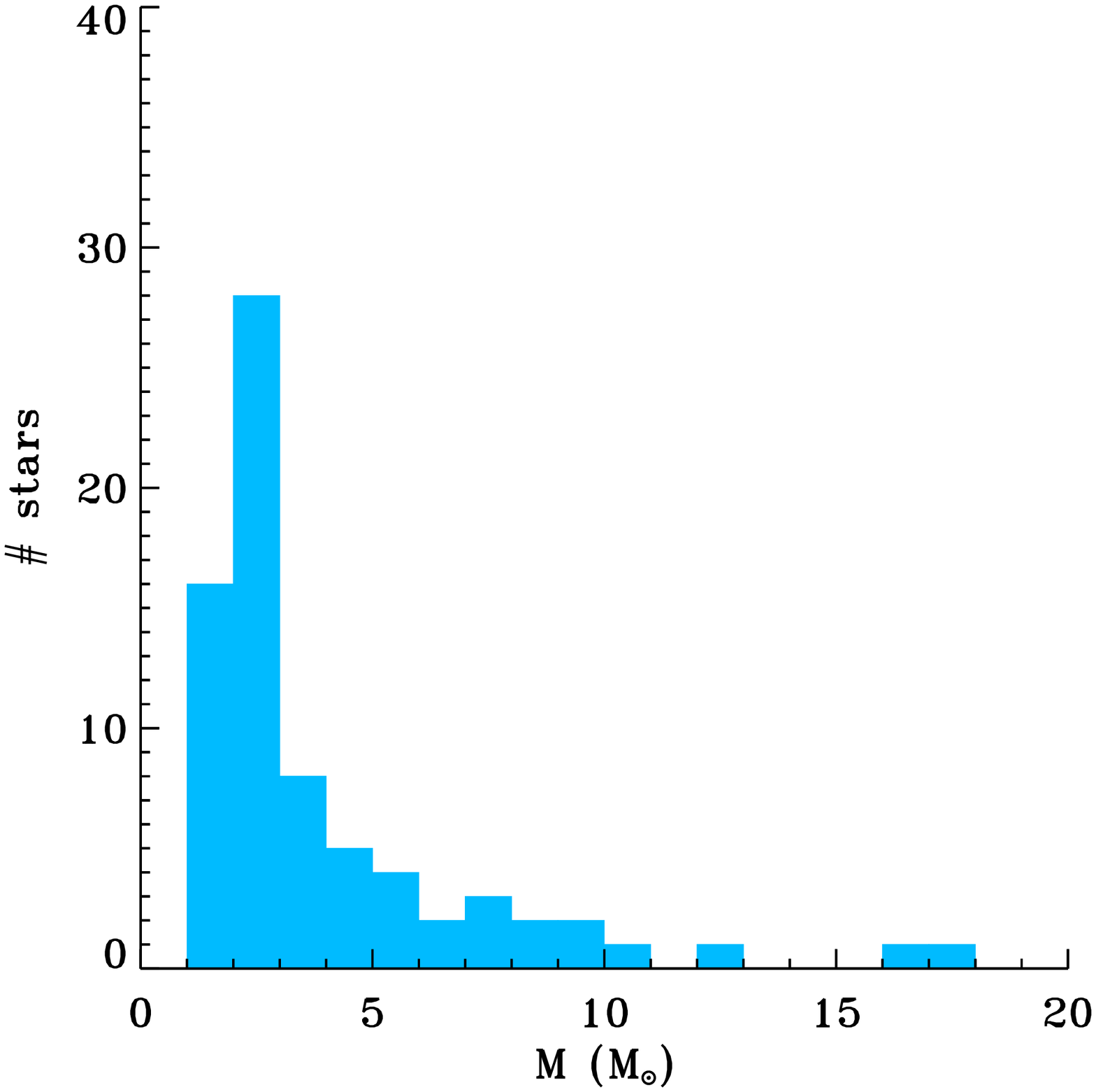}
\hfill
\includegraphics[width=7cm]{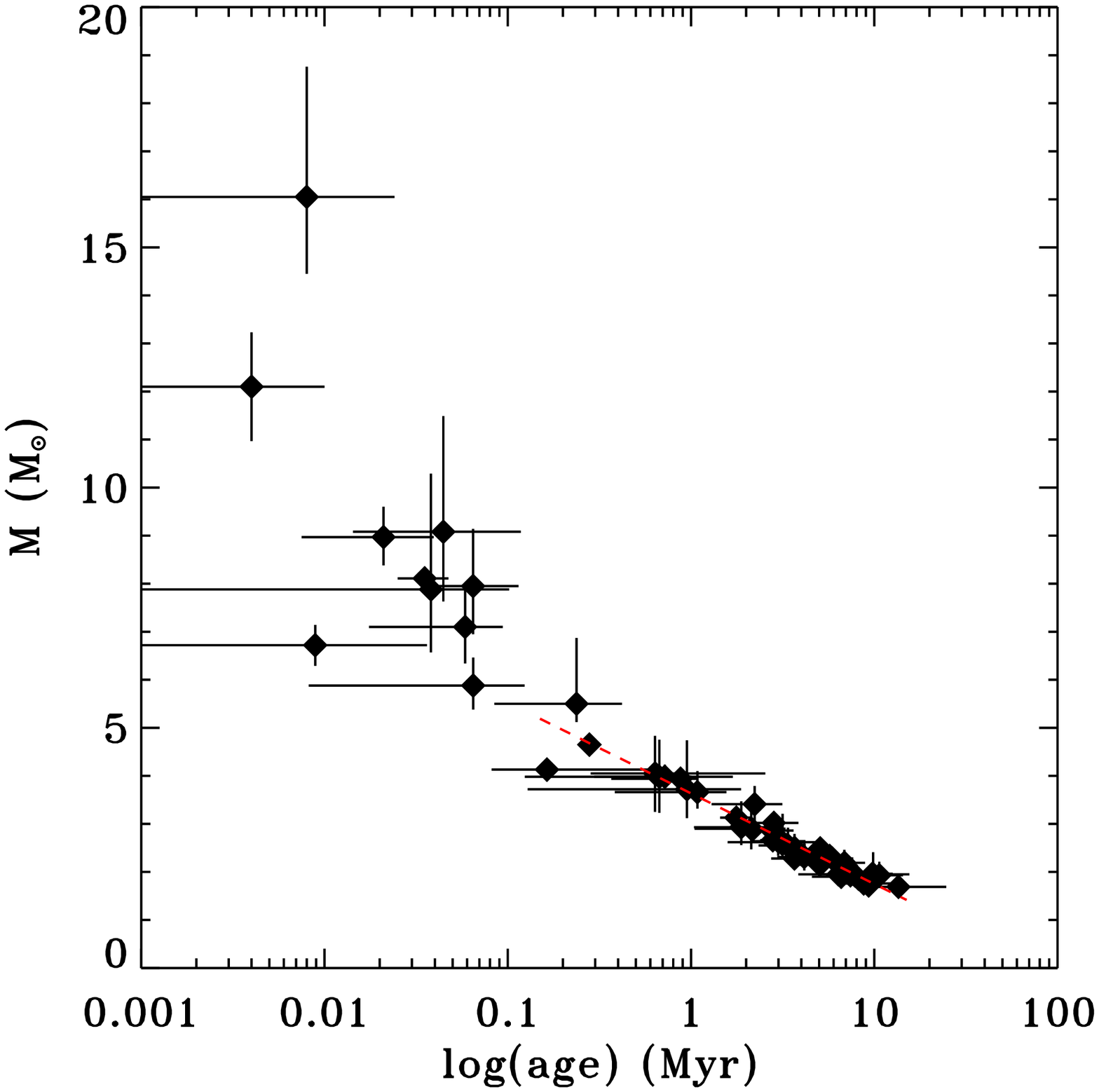}
\caption{{\it left}: Mass distribution of the sample. {\it right}: Masses of the stars as a function of their age. { The red dashed line is a fit to the data points for masses lower than 5 \msun}.}
\label{fig:mas}
\end{figure}

Our sample contains { 70 field Herbig Ae/Be \footnote{few of them are now known to belong to associations such as Orion, but were thought to be field stars at the beginning of this study}}. In paper I we used the photometric data and distances of the stars, compiled from the literature, to give an estimation of their luminosity. By fitting our observations with synthetic spectra, we measured their effective temperature. We then placed the stars in an HR diagram, and compared their position with PMS evolutionary tracks computed using the CESAM code, version 2K \citep{morel97} with solar abundances, in order to derive the mass, radius and age of each star, as well as the expected radius of the star when it reaches the ZAMS (Table 2 of paper I). { For one star, HD 34282, that has a low metallicity compared to the sun \citep{merin04}, we used the age, mass and radius determination of \citet{merin04}, however we were not able to estimate the ZAMS radius, because our CESAM models are of solar metallicity. Therefore we have rejected this star in the following analyses}. More details on these procedures are given in Paper I. In Fig. \ref{fig:hr} the stars are plotted in a HR diagram superimposed on PMS evolutionary tracks and isochrones. The mass distribution of our sample, as well as the masses as a function of ages, are plotted in Fig. \ref{fig:mas}. For two stars (HD 98922 and IL Cep) that fall outside of the PMS region of the HR diagram, we couldn't estimate their mass, radius and age. Therefore they won't be included in the following analyses.

In Fig. \ref{fig:hr} we include the birthline as computed by \citet[][BM01 hereinafter]{behrend01}. For reasons explained in Paper I, we favoured their birthline to those of \citet{palla93}. The birthline of BM01 has been computed by assuming a modulated accretion rate during the proto-stellar phase that increases as the star gains in mass. Once the close environment becomes poor in gas, the strong accretion phase stops and the star evolves on the PMS track from the birthline to the zero-age main-sequence (ZAMS). The age of each star has been computed from the birthline. This age indicates therefore the time that the star spent on the pre-main sequence phase, and does not include the time of the formation of the star. As a result some stars, situated in the HR diagram very close to the birthline, can have ages as young as few thousands of years.

We have plotted in Fig. \ref{fig:hr} the limit after which the mass of the convective envelope is less than 1\% of the mass of the star. This limit represents the end of the convective phase and the start of the radiative phase of PMS evolution, during which the star is predominantly radiative (a small convective core appears only at the very end of the PMS phase, when the luminosity starts to decrease again). We observe in Fig. \ref{fig:hr} that the less massive stars of the sample are more concentrated toward the ZAMS and that the contrary is observed at high mass, as already reported by \citet{vieira03}. We suggest that this is due to a selection effect, using Herbig Ae/Be criteria. At high mass the close stellar environment dissipates faster than at low mass, due to strong winds and photoionisation from the stellar UV flux \citep[e.g.][]{meeus01,dullemond01,alonso09}. Therefore, if stars are selected using Herbig Ae/Be criteria that probe a CS environment rich in dust and gas, we are more likely to select stars at earlier PMS evolutionary stages in massive stars than in lower mass stars, before the the CS environment has dissipated sufficiently to become undetectable in the IR or in the spectrum. We argue that PMS massive stars close to the ZAMS could have already dissipated their CS environment. As a result they do not show anymore Herbig Ae/Be criteria, and therefore cannot be part of our sample.

In the case of low-mass stars, the whole sample is concentrated in the radiative phase of PMS evolution, { as illustrated by the dearth of points on the right of the convective/radiative phase transition in Fig. 1. This part of the HR diagram is in fact populated with the intermediate-mass T Tauri stars (IMTTS). The IMTTS are the evolutionary progenitors of the Herbig Ae/Be stars with spectral types ranging from K to F \citep[e.g.][]{calvet04},} which explains why they have not been included in our sample. All these reasons explain why a strong correlation is found between the mass and the age of the stars of our sample.

{ In Fig. 2 (bottom), the age-mass correlation is evident for masses lower than 5 \msun. However, between 5 and 10 \msun, while the stars are younger than at lower mass - which is expected as above 5 \msun\ the PMS lifetime is lower than 0.4 Myr - the ages are highly scattered and no correlation of the ages at these masses is evident. In order to estimate the effect of the correlation found at lower mass on the analyses described in Section 3, we have performed a linear fit of the data points. We find a relation: $M/{\rm M}_{\odot} = 3.6 - 1.9\log (t/1 {\rm Myr})$, where $t$ is the age. }

%
%
%________________________________________________________________

\section{Analysis of the projected angular momenta of the sample}

One of the aims of this survey is to understand angular momentum evolution during the PMS phase of intermediate mass stars, and to estimate the role of the magnetic field in this evolution. We present below two different analyses to address these questions. The first aims to find a correlation of the angular momentum of a sample of HAeBe stars with their age, while the second compares the \vsinis distributions of the same HAeBe sample to various samples of MS A/B stars in order to investigate if angular momentum evolution is expected in the future of HAeBe stars, before they reach the ZAMS (Section 4).

Today, we have no knowledge of the internal rotation profile of Herbig Ae/Be stars. During its evolution, the rotation profile of a PMS star is expected to change as the star contracts and as angular momentum is transported through various mechanisms not fully understood yet. While understanding the behaviour of internal rotation in PMS stars is an important field of theoretical research, it is also becoming observationally accessible thanks to the CoRoT and Kepler satellites.  However, our knowledge of internal rotation on the PMS is still very poor. We therefore can consider two extreme hypotheses: ({\it i}) a solid body rotation (SR), and ({\it ii}) a { constant specific (i.e. per unit of mass) angular momentum (CS), between which the real internal rotation profile would be. The solid body rotation (hypothesis {\it i}) assumes that the transport of angular momentum inside the star is highly efficient all along the PMS evolution. This hypothesis can be considered as reasonable inside a fully convective star, such as low-mass stars at the beginning of the PMS phase. However at intermediate-mass, during the major part of their PMS evolution, the stars are mainly radiative. Hypothesis ({\it i}) is therefore too far from the reality for Herbig Ae/Be stars. On the contrary, if we consider an isolated star without exchange (of matter or angular momentum) with its environment and without internal transport of angular momentum, we can assume a constant specific angular momentum. As the star contracts, the mass will get more and more concentrated into the core and the angular momentum will tend to behave as $1/r^2$. In a radiative interior we would expect the transport of angular momentum to be much less efficient than in a convective interior and therefore hypothesis ({\it ii}) would be more reasonable for the radiative phase of the PMS evolution. However, in reality during the PMS phase a star will exchange matter, and therefore angular momentum with its environment. Internal transport of angular momentum through various processes, such as pulsations or meridional circulation, are also expected to be active with an efficiency that is unknown. The real internal profile is therefore expected to lie between both hypotheses.}

We have calculated the angular momentum of each star in our sample using the internal density profile at the previously estimated age, using CESAM models, and both assumptions on the stellar rotation profile. In the case of a spherically symmetric star, the angular momentum element $dJ$ is given by: 
\begin{equation}
dJ = \frac{2}{3}r^2\omega(r)dm,
\end{equation}
where $dm$ is the mass of a spherical shell inside the star at a distance $r$ from the center, and $\omega(r)$ is the angular velocity of the shell.

In the case of solid body rotation, the angular velocity is constant as a function of $r$ and equal to the angular velocity at the surface of the star: $\omega(r) = \Omega_*, \, \forall r \le R_*$. The total stellar angular momentum is therefore computed by integrating the following equation throughout the volume of the star:
\begin{equation}
J_{\rm SR} = \frac{8\pi}{3}\Omega_* \int\rho(r) r^4dr,
\end{equation}
where $\rho(r)$ is the density of the star as a function of radius.

In the second case, the specific angular momentum $dJ/dm$ is constant and equal to $\frac{2}{3} R_*^2\Omega_*$ at the stellar surface. The angular velocity is therefore dependent on $r$ and equal to:
\begin{equation}
\omega(r) = \left(\frac{R_*}{r}\right)^2\Omega_*,
\end{equation}
and the angular momentum is computed as follows:
\begin{equation}
J_{\rm CS} = \frac{8\pi}{3}R_*^2\Omega_*\int\rho(r)r^2dr.
\end{equation}

The angular velocity at the surface of the star is computed using the estimated radius $R_*$ and the projected rotational velocity \vsini. As $\sin i$ is unknown, the angular velocities, hence the angular momenta, of the stars are computed up to a factor of $\sin i$. We assume that whatever the ages, masses, or situation within the Galaxy of large samples of stars, the distributions of $\sin i$ can be considered similar in all cases. We therefore expect that if all quantities considered in one statistical analysis are computed or measured up to a factor of $\sin i$, it won't affect the result.

\begin{figure}
\centering
\includegraphics[width=8cm]{amsr_age9.ps}
\includegraphics[width=8cm]{logamsr_age9.ps}
\caption{Angular momentum of the sample as a function of age, assuming a solid body rotation { with a normal y-axis scale (upper) and a logarithmic y-axis scale (bottom)}. The circles represent the normal stars, while the squares represent the magnetic stars. The symbols are defined as follows: large symbol: $M > 10$~M$_{\odot}$, medium filled symbol: $7 < M < 10$~M$_{\odot}$, medium open symbol: $5 < M < 7$~M$_{\odot}$, small symbol: $M < 5$~M$_{\odot}$. { The black lines plotted in the bottom panel are the predicted $J_{\rm SR}$ of a sample of stars evolving at constant angular momentum, with a mass-age correlation similar to that of our sample (see Section 2), and with vsini of 10 \kms\ (dashed), 50 \kms\ (dot-dot-dot-dashed), 100 \kms\ (long-dashed), and 250 \kms (dot-dashed). The bumps observed around 1~Myr mark the transition between fully radiative stars (on the left of the bump), and stars with a convective core (on the right). The upper or lower error bars in $J$  that appear very small, are for stars very close to the birthline or the ZAMS.}}
\label{fig:amage}
\end{figure}

We have divided our sample into two distinct sub-samples, one containing the magnetic HAeBe stars and the other containing the non-magnetic, non-binary stars (or "normal stars"). The magnetic sample contains the five confirmed magnetic HAeBe stars (HD 200775, V380 Ori, HD 190073, HD 72106 and LP Ori) that have been reported in paper I. The normal sample contains the 58 stars in which no magnetic fields have been detected and which are not member of a close binary system (paper I). { From this sample we exclude MWC~1080 that has no \vsini\, HD 250550 that has no estimation of its fundamental parameters, and HD 34282, that has no estimation of a ZAMS radius. The final normal sample considered in this study contains 55 stars}. Both magnetic and normal samples will be compared, then treated separately, as the magnetic fields are expected to have a strong influence on the rotational evolution of the stars. The non-magnetic binary stars have not been included into the analysis as tidal interaction between both components is expected to modify the stellar angular momentum.

%________________________________________________________________

\subsection{The magnetic stars}

The favoured hypothesis explaining the slow rotation of magnetic Ap/Bp stars is magnetic braking occurring during the PMS.  This would slow magnetic stars much more than non-magnetic stars, through the interaction of the stellar magnetic field with its environment \citep{stepien00}. According to this hypothesis a change of the angular momentum of PMS A/B stars is expected between magnetic and non-magnetic stars. To test this hypothesis we have plotted, in Fig. \ref{fig:amage}, the solid body rotation angular momentum ($J_{\rm SR}$) as a function of age for the magnetic (squares) and normal (open and filled circles) HAeBe stars. The size of the symbol is indicative of the stellar mass. We don't observe any correlation of the angular momentum of the magnetic stars with age. The same is observed when a constant specific angular momentum is assumed instead. However, when considering the small number of magnetic stars in this sample, the range of stellar mass covered by this sample, and the fact that the angular momenta have been calculated up to a factor $\sin i$, if a correlation exists, it would be difficult to detect. 

{ On the other hand, even if the sample is small, it is worth noting that four out of the five magnetic stars have an angular momentum much lower than the normal sample. The same is observed in the case of a constant specific angular momentum.}

%________________________________________________________________

\subsection{The non-magnetic stars}

Even in the absence of strong magnetic fields, HAeBe stars can experience an exchange of matter with their environment through accretion disks and/or winds, which are observable through the emission properties of their spectra. Therefore the stars can exchange angular momentum with their environment. Furthermore, while the magnetic stars are those with strong detected magnetic fields, there is still the possibility that, in the sample of the "non-magnetic" stars, some stars could host magnetic fields too faint to be detected (see Paper III). If such fields exist they could enhance the loss of angular momentum through an exchange of matter occurring during the PMS phase of their evolution. We therefore performed an analysis of the angular momentum of the non-magnetic stars in order to understand their angular momentum evolution during the PMS phase.

In Fig. \ref{fig:amage}, while the angular momenta of the normal stars generally spread from $J_{\rm SR} = 0$ to a maximum value, which is due to the uncorrected $\sin i$ factor, the maximum value clearly decreases with age { whatever the mass. In the non logarithmic plot (Fig. 5 top)}, this trend is most significant in the middle sub-sample with masses between 5 and 10 M$_{\odot}$ (medium open and filled circles). We checked that at these masses the trend is not due to the age-mass correlation pointed out in Section 2 by observing that the graph area covered by a 5-7 M$_{\odot}$ sub-sample (open medium circles) is approximately the same as the area covered by a 7-10 M$_{\odot}$ sub-sample (filled medium circle). Furthermore, between 5 and 10~M$_{\odot}$, { as pointed out in Section 2, no correlation of the age and the mass is observed.

In the logarithmic plot of the angular momentum (Fig. 3 bottom), a clear decrease of the angular momentum of the stars with masses below 5 \msun\ (small open circles) is observed. However at these masses, a strong age-mass correlation is present in our sample. In order to check if it can explain this decrease, we have calculated the expected angular momentum of a simulated sample of stars of various masses between 1.5 and 5 \msun, evolving at constant angular momentum, with ages following the correlation found in Sec. 2, and for four \vsini: 10, 50, 100 and 250 \kms. We observe (Fig. 3 bottom) that the trend is well reproduced by the angular momentum of the simulated sample, and that most of the lower mass stars lies between the curves of 50 and 250 \kms\ in \vsini, values that are consistent with the \vsini\ of most of our stars. Therefore, the age-mass correlation of our sample at lower mass can explain alone the decrease of the angular momentum observed in Fig. 3 (bottom). 
}

In order to test the impact of the choice of the internal rotation profile, we did the same plots using the total stellar angular momentum computed by assuming a constant specific angular momentum. While the total angular momentum of the stars is generally found to be larger than in the case of a solid body rotation, we obtain a similar plot, with the same general trend of the maximum values decreasing with time for stellar masses above 5~M$_{\odot}$. { At lower mass, a trend similar to the SR case is observed and is also fully explained with the age-mass correlation that our sample shows at these masses}. We therefore find that, whatever the stellar internal rotation profile considered, the highest mass ($M > 5$~M$_{\odot}$) PMS stars seem to loose angular momentum as they evolve towards the ZAMS, while the lower mass stars seem to evolve with constant angular momentum.

%\begin{figure*}
%\centering
%\includegraphics[width=6cm,angle=90]{amsr_age07_mbin.ps}
%\caption{Angular momentum of the stars plotted as a function of age for 3 distinct mass bins. In the right panel the meaning of the symbols is as follows: black large circle : $M > 9$ M$_{\odot}$, medium blue circle: $5 < M < 9$ M$_{\odot}$, and small red circle: $3.4 < M < 5$ M$_{\odot}$.}
%\label{fig:amsr}
%\end{figure*}

%
%
%________________________________________________________________

\section{Analysis of the projected rotational velocities}

\begin{figure}
\centering
\includegraphics[width=8.5cm]{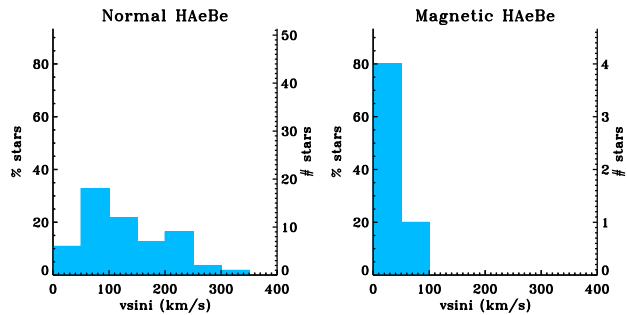}
\caption{\vsinis histograms of the magnetic (left) and the non-magnetic HAeBe stars.}
\label{fig:statmg}
\end{figure}

\subsection{The magnetic stars}

A strong dichotomy is observed in the rotation velocities of non-magnetic and magnetic MS A/B \citep[e.g.][]{abt95}: the former are fast rotators, while the latter are much slower rotators (with rotational periods generally longer than one day). In order to understand this dichotomy and its evolution, we have plotted, in Fig. \ref{fig:statmg}, the \vsinis distributions of the magnetic ({\it right}) and non-magnetic ({\it left}) HAeBe stars. We observe that already during the PMS phase there is a dichotomy similar to that observed among MS A/B stars. All magnetic HAeBe stars rotate with \vsini\ lower than 100 \kms. Since the number of magnetic stars is small, one could argue that the low \vsini\ seen on these stars could be due to a low inclination of the rotation axis. However, the magnetic analysis of the individual stars allows for a direct determination of their rotation periods, independent of the inclination. We find periods ranging from 1.6 to 4.3 days \citep{alecian08a,folsom08,alecian09b}. We can therefore conclude confidently that all magnetic HAeBe stars of our sample are slow rotators, while only part of the non-magnetic stars of our sample rotate slowly (or have a low \vsini). 

We performed a p-value statistical test to check the hypothesis that the \vsini\ of the magnetic sample can exist by chance, if we assume that the \vsini\ of the magnetic and normal sample follow the same distribution. In this aim we have computed the Student's $t$-value and the associated significance p-value \citep{press92}. We find a p-value of 0.002, that is much lower than 0.05, meaning that the null hypothesis - that the observed \vsini\ values of the magnetic stars can be ascribed to chance alone - is rejected. The differences between both distributions are therefore real and have a physical meaning.

%________________________________________________________________

\subsection{The non-magnetic stars}

The rotational velocities of MS A/B stars have already been largely measured \citep*[see][and references therein]{royer07}, and their global distribution is known. We can therefore search for a potential evolution in the angular momentum during the PMS by predicting the rotation velocities that HAeBe stars will have once they will reach the ZAMS and comparing them to the distribution of MS \vsini. There are many reasons to believe that the evolution of angular momentum in A/B stars is negligible during the MS \citep[e.g.][]{stepien00}. Furthermore, { as the evolution of a MS star is slower during the first half of the MS, most MS stars are closer to the ZAMS than the terminal age main sequence (TAMS). Hence, it is reasonable to assume that the stars from the MS catalogues discussed below can be treated approximately as ZAMS stars.} The ZAMS $v\sin i$, that we have computed, can therefore be directly compared to the \vsinis of general MS stars.
%, comparisons that have been done in many ways, and are described below.

Assuming a uniform angular momentum evolution through the PMS, we have estimated the rotational velocities on the ZAMS for our sample as follows:
\begin{equation}
v_{\rm ZAMS}\sin i = v\sin i \frac{R_{\rm ZAMS}}{R_*}\frac{I_*}{I_{\rm ZAMS}}
\end{equation}
where, $I_*$ and $I_{\rm ZAMS}$ are the moment of inertia at the current stellar age, and on the ZAMS, respectively. The moments of inertia have been computed by integrating the following equations for both stellar rotation profile assumptions:
\begin{equation}
I_{\rm SR} = \frac{8\pi}{3}\int \rho(r) r^4 dr
\end{equation}
\begin{equation}
I_{\rm CS} = \frac{8\pi}{3}R_*^2\int \rho(r) r^2 dr
\end{equation}
For both angular momentum treatments, the resulting distributions of ZAMS \vsini\ show clear differences when compared to the PMS \vsini\ one: they are shifted towards larger \vsini\ and are more spread. It is the result of a decreasing of the radius and an increase of the density.

We have considered different samples of rotational velocities of MS stars to compare our results with. The following describes in detail the comparisons with both angular momentum treatments.

\begin{figure*}
\centering
\includegraphics[width=6.5cm,angle=90]{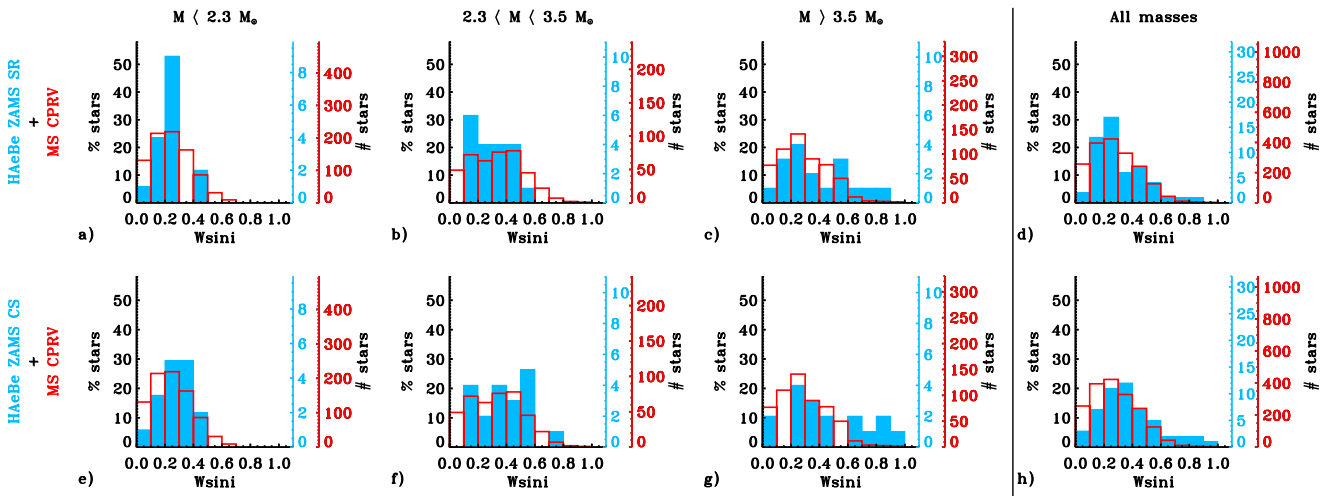}
\caption{\wsini\ histograms of the HAeBe stars projected on the ZAMS (blue field bares) assuming a solid body rotation (top) or assuming a constant specific angular momentum (bottom), and of the MS stars from the catalogue CPRV (red empty bares) \citep{glebocki00}, for three different mass ranges: $M<2.3$~M$_{\odot}$, $2.3<M<3.5$~M$_{\odot}$, $M>3.5$~M$_{\odot}$.}
\label{cprv}
\end{figure*}

%________________________________________________________________

\subsubsection{Comparison with the CPRV catalogue}

We used the Catalogue of Projected Rotational Velocities (CPRV) of \citet{glebocki00} which is a compilation of the observed \vsinis of over 11000 stars of all spectral types and luminosity classes. This catalogue includes numerous classes of stars that should not be taken into account in our analysis for reasons explained below. We have selected the suitable stars as follows:
\begin{enumerate}
\item we excluded all stars with uncertain \vsini\ { (as indicated in the catalogue)}: the number of stars in the catalogue is sufficiently large to keep only well constrained \vsini.
\item when multiple \vsinis have been published for a star, we kept the more accurate value { (i.e. the one with the smallest error bare)}.
\item we only kept stars with luminosity class V which can be compared to the ZAMS \vsinis of HAeBe stars.
\item we excluded binary systems, as identified in the CPRV: tidal interactions between the components of binary systems lead to a strong rotational braking. These stars should therefore not be taken into account in our test of angular momentum conservation.
\item we exclude all peculiar stars: all Ap/Bp stars are magnetic and should be rejected, as we are considering only non-magnetic stars in this section. Among the other class of peculiar stars, Am stars exhibit a binary incidence close to 100\%; for reasons explained above, we exclude them too.
\item we only kept intermediate mass stars: in order to determine the masses of MS stars, we used the Stellar Mass Catalogue (SMaC) of \citet{belikov95}, that contains the dynamical determination of the masses of binary systems of luminosity class V, as well as their spectral types. By plotting $\log(M/M_{\odot})$ as a function of spectral type, we found two linear regressions, one for spectral types earlier than A0 and one for spectral types later than A0 (cf. Appendix A). Using these calibrations and the tabulated spectral type in the CPRV, we determined the masses of the stars. Only masses between 1.5 and 20 M$_{\odot}$ have been used.
\end{enumerate}

Some binary systems are indicated in the CPRV, but in order to exclude all of them, we used the most recent version of the Washington Double Stars (WDS) catalogue\footnote{The catalogue has been downloaded from the WDS website: http://ad.usno.navy.mil/wds/, and the last update was Oct. 2$^{\rm nd}$ 2008} of \citet{mason01}. We first rejected from the WDS catalogue, all dubious binaries or multiple systems and all non-physical systems, as indicated in the catalogue itself. Then using the coordinates of the CPRV and the WDS catalogues, we calculated the distances between the stars of both catalogues. Whenever the distance was lower than 90 arcsec (which is roughly the precision on the coordinates of the CPRV stars) the star was considered as double and was rejected. The final number of stars of the CPRV suitable for our analysis is 1823 stars.

The mass range of our HAeBe sample is broad (from 1.5 to 17~M$_{\odot}$, see Table 2 of Paper I). Many phenomena during the PMS phase are strongly sensitive to the mass (e.g. stellar winds, accretion disks, PMS lifetime). Therefore we divided the ZAMS HAeBe and MS samples into three sub-samples of different mass ranges, in order to keep the number of stars equal in each bin:  
\begin{itemize}
\item $\#$1: $M<2.3$~M$_{\odot}$, 
\item $\#$2: $2.3<M<3.5$~M$_{\odot}$, and 
\item $\#$3: $M>3.5$~M$_{\odot}$.
\end{itemize}
Each one of the ZAMS HAeBe sub-samples contains 19, 19 and 17 stars respectively, while the MS CPRV sub-samples contain 853, 415 and 567 stars, respectively. { We first compared the \vsinis distributions of the ZAMS HAeBe and the MS stars for the three mass ranges. However, while the two first sub-samples cover a relatively low-mass range (compared to the typical error bars, see Fig. 1), the third sub-sample covers a wide mass range that need to be taken into account when comparing two \vsini\ distributions. The mass distributions of the HAeBe and CPRV samples are different: a lack of high-mass stars (above 9~\msun) and an excess of lower mass stars (below 5~\msun) is observed in the HAeBe distribution with respect to the CPRV one. Due to these differences, comparing two \vsini\ distributions is meaningless as the potential differences could be due to either a difference in mass (or radius) or in angular momentum. Therefore, in order to minimise this effect, we have chosen to compare the rotation rates (\wsini), instead of the \vsini, which are defined as the ratios of the \vsini\ to the break-up velocity:
\begin{equation}
W\sin i = \frac{v\sin i}{(GM/R_{\rm ZAMS})^{1/2}}
\end{equation}
where $M$ and $R$ are the mass and radius of the star, and $G$ is the gravitational constant. The masses of the HAeBe sample have been determined in paper I. The mass of the CPRV sample have been estimated by using a mass-radius relation. This relation has been determined by using the Catalogue of orbital elements, mass and luminosities of close double stars of \citet{svechnikov84}. In this aim we first performed a selection on the stellar parameters as follows:
\begin {itemize}
\item we kept only stars of luminosity class V
\item we excluded all stars with uncertain or unknown mass ratio, mass, or radius
\item we kept stars more masive than 1~\msun.
\end{itemize}
Then we plotted the $\log(R/R_{\odot})$ as a function of $\log(M/M_{\odot})$ and performed a linear regression (see Appendix B).

In Fig. \ref{cprv} are plotted the distributions of the rotation rates \wsini\ of the ZAMS HAeBe and the MS CPRV stars for all masses, for the three mass ranges, and for both angular momentum assumptions (SR and CS). Whatever the mass, the ZAMS SR and the CPRV look similar. However, when comparing the ZAMS CS and CPRV distributions, all masses together, we note some differences: the first one looks slightly shifted towards larger \wsini, and it shows a lack of slow rotators in favour of fast rotators. These differences seem to be reproduced in the third mass sub-sample.

In order to determine if the differences that our eye seem to note are significant, we performed two statistical tests. Both tests estimate the significance level of the rejection of the null hypothesis (i.e. both distributions are the same). First, we performed a Kolmogorov-Smirnov (KS) test as described by \citet{press92}. This test compares the cumulative distributions of both samples. It is sensitive to the shift of one distribution with respect to the other. 
{ When we apply it to each panel of Fig. 5, the significance levels are all below 95 \% except for the panels g (99\%) and h (98\%). It confirms that real differences exist between the ZAMS CS and the CPRV catalogue at high mass ($M>3.5$\msun). In the case of a solid body rotation, the highest significance level is at high-mass, with a value of 93\%. However we consider that 93\% is not statistically significant (typical values would be between 95 and 99\%).}

The second test that we performed is a chi-square test, as described by \citet{press92}. This test compares two distributions bin per bin by calculating the chi-square between both distributions. By performing a standard chi-square test we estimate the probability of rejection of the null hypothesis. { As with the KS test, we find that the ZAMS HAeBe distributions in the panel h only is significantly different to the CPRV distribution at a level of 99.993\%. In the case of a solid rotation, the highest probability of rejection is found at high mass (panel c) with a value of 87\%, which is not enough to confidently reject the null hypothesis.

Both tests find that the differences between the ZAMS CS and the CPRV catalogue are significant at masses above 3.5~\msun, indicating that we predict too much fast rotators by assuming a constant angular momentum. In the case of solid body rotation, above 3.5~\msun, it also seems that an excess of fast rotators is predicted, compared to MS stars. However neither statistical test is able to quantitatively confirm this}

\subsubsection{Comparison with the RZG catalogue}

In order to verify that our above conclusions are not sample-dependent, we performed the same test with the catalogue of rotational velocities of A-type stars of \citet*[][RZG hereinafter]{royer07}. We rejected all the binaries and chemically peculiar stars from this sample, as indicated in the catalogue itself, but also as indicated in the WDS catalogue. The final number of stars from this catalogue, considered in this study, is 981. The mass of the stars in the RZG catalogue are all below 2.6~M$_{\odot}$. We therefore divided both ZAMS HAeBe and RZG samples in two sub-samples: $M<2.15$~M$_{\odot}$ and $2.15<M<2.6$~M$_{\odot}$. The numbers of stars are 13 in each of the ZAMS HAeBe sub-samples, and they are 496 and 485 in the RZG sub-samples, respectively. The distributions are plotted in Fig. \ref{fig:rzg}. In spite of some irregular ZAMS HAeBe distributions, we don't detect with the eye any systematic differences between the distributions. The KS test gives significance levels below 95\% for all panels. Therefore, whatever the treatment of the internal angular momentum, no difference is found between the RZG and ZAMS HAeBe samples.

\begin{figure}
\centering
\includegraphics[width=6.5cm,angle=90]{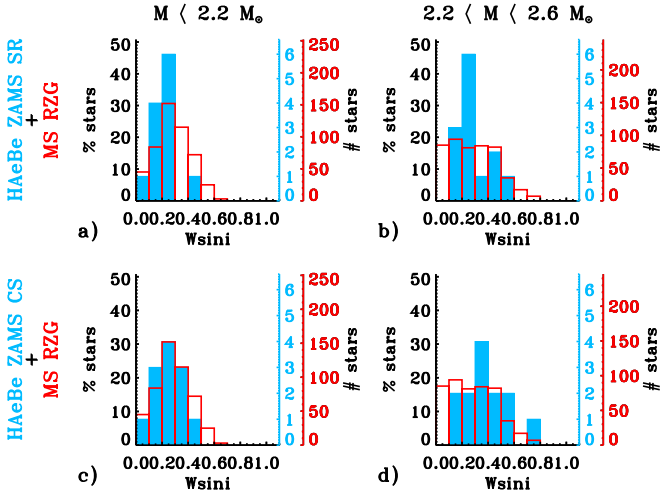}
\caption{\wsini\ histograms of the HAeBe projected on the ZAMS (blue field bares) assuming a solid body rotation (top) or assuming a constant specific angular momentum (bottom), and of the MS stars from the catalogue RZG (red empty bares) \citep{royer07}, for two different mass ranges: $M<2.15$~M$_{\odot}$, $2.15<M<2.6$~M$_{\odot}$.}
\label{fig:rzg}
\end{figure}

%________________________________________________________________

\subsubsection{Comparison with the $\alpha$~Per open cluster}

Finally, we performed a third comparison with the $W\sin i$ distributions of the intermediate mass stars in the open cluster $\alpha$~Per, from the catalogue of \citet{prosser92}. These stars are co-evolved and the age of the cluster is estimated to be around 80~Myr. The duration of the PMS phase for intermediate mass stars is 30~Myr maximum (see Table 2 of Paper I). The $\alpha$~Per sample is therefore more representative of intermediate mass ZAMS stars than the other samples used above. From the original $\alpha$~Per catalogue we performed the following selections:
\begin{enumerate}
\item we excluded all stars with uncertain $v\sin i$.
\item we excluded binary systems, using the WDS catalogue.
\item we only kept intermediate mass stars: we used the same method as the CPRV catalogue to determine the masses. This method requires a known spectral type. If the spectral type is unknown in the catalogue of \citet{prosser92}, we used the one indicated in the SIMBAD Astronomical database \footnote{http://simbad.u-strasbg.fr/simbad/}.
\end{enumerate}

We divided both the ZAMS HAeBe and $\alpha$~Per samples into 3 sub-samples of the same mass ranges as the CPRV ones. The numbers of stars in the $\alpha$~Per sub-samples are 30, 20, and 9, respectively. Below 10 stars, we consider that the number of stars is too low to perform a reliable statistical analysis. We therefore only consider the two first mass ranges: $M<2.3$~M$_{\odot}$ and $2.3<M<3.5$~M$_{\odot}$. The distributions are plotted in Fig. \ref{fig:aper}. The ZAMS and MS distributions are highly irregular, imputable to small samples. In spite of that panel b) seems to show some differences: a lack of fast rotators and an excess of slow rotators is predicted in the ZAMS SR sample. The KS test gives significance levels below 95~\% in all panels except in panel b, where it finds 96~\%, in contradiction with the results obtained with the CPRV and RZG samples.

\begin{figure}
\centering
\includegraphics[width=6.5cm,angle=90]{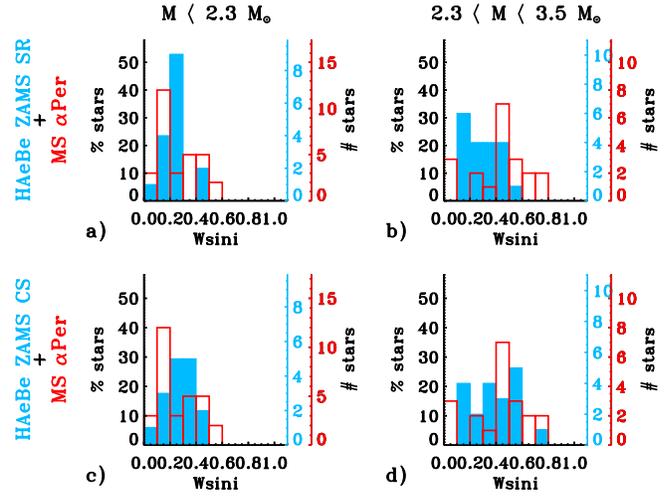}
\caption{\wsini\ histograms of the HAeBe projected on the ZAMS (blue field bares) assuming a solid body rotation (top) or assuming a constant specific angular momentum (bottom), and of the MS stars from the catalogue $\alpha$Per (red empty bares) of \citet{prosser92}, for two different mass ranges: $M<2.3$~M$_{\odot}$, $2.3<M<3.5$~M$_{\odot}$.}
\label{fig:aper}
\end{figure}

%
%
%________________________________________________________________

\section{Conclusions and Discussion}

This paper is the second in a series that presents the results of a high-resolution spectropolarimetric analysis of a sample of 70 Herbig Ae/Be stars. While paper I describes the data and paper III will examine the magnetism of Herbig Ae/Be stars, here we address the problem of stellar angular momentum evolution during the pre-main sequence phase at intermediate mass.

Using the \vsinis values published in paper I, we performed two separate analyses. In both analyses we have used two internal rotation profile assumptions: a solid rotation, and a constant specific angular momentum. As the internal rotation profiles of Herbig Ae/Be stars is unknown, we chose these two extreme hypotheses for our analysis. In the first analysis we have computed the angular momentum of each star of our sample to search for a correlation with the age for the magnetic sample on one side, and the normal (i.e. non-magnetic and non-binary) sample on the other side. In the second analysis, we first compared the \vsinis distributions of the magnetic to the normal samples, then we predicted the \vsinis that our normal sample will have on the ZAMS using both above assumptions, and compared the derived distributions to published \vsinis distributions of intermediate mass MS stars. In both analyses we have analysed the normal sample within mass ranges, in order to avoid mass-dependent conclusions. { However, in order to minimise the differences found between both ZAMS and MS mass distributions above 3.4~\msun, we have derived the rotation rates \wsini\ (i.e. the \vsini\ over the escape velocity) of the samples, and compared the \wsini\ distributions (instead of the \vsini). }

%________________________________________________________________

\subsection{Magnetic field impact on the angular momentum evolution}

Both analyses revealed that the magnetic Herbig Ae/Be stars { of our sample} are much slower rotators than the normal ones, as observed on the main sequence, suggesting that the magnetic stars have experienced more braking than the normal stars. Table 2 of paper I provides the ages of the stars and the duration of their PMS phases. We observe that, while some of the magnetic stars have completed a large fraction of their PMS phase, one of them (HD 190073) has only completed a third of its PMS phase. Furthermore, HD 200775 has only an age of 16\,000 years. These stars are very young and are already rotating very slowly. If magnetic braking is responsible for these slow rotators, it must act very early, and relatively fast (in less than 16\,000 years), during the PMS phase, or even before the PMS phase.

We could propose an alternative to the magnetic braking theory. We observe, in a large number of HAeBe stars, wind signatures in their UV and optical spectra \citep[e.g. ][paper III]{finkenzeller84}. Winds seem therefore frequent during the PMS phase at intermediate mass. A very young star, newly formed, and starting its PMS phase is rotating very fast and experiences mass loss, and therefore angular momentum loss, due to its wind. The surface of the star is therefore rotating more slowly than the core. A velocity gradient appears between the surface and the core, inducing sheared turbulence \citep{lignieres96}. We argue that this turbulence could be sufficiently strong - depending on the rotational energy compared to the magnetic energy - to dissipate the magnetic fields inside very young stars \citep[][]{auriere07}. Therefore, only relatively slow rotators could keep their magnetic fields. However this theory requires that a population of slowly rotating stars exists before they start their PMS evolution.

Our analysis does not allow us to favour either of these hypotheses for the following reasons. First the number of known magnetic HAeBe stars is very small. It is therefore difficult to perform any statistical analysis that could reveal a dependance of their angular momentum with magnetic strength, age, environment, initial conditions, or stellar properties (e.g. masses or rotation period). Secondly, we need to better characterise the time-scales of the magnetic braking, or magnetic dissipation which could help us to favour one of the hypotheses, when compared to the ages of the stars. Third, the uncertainties on ages, initial conditions and environment are very large among field HAeBe stars. It is necessary to perform the same analysis on HAeBe star members of very young clusters of different well determined ages, in which sets of stars have the same age and same initial conditions. Such a study could help us to disentangle evolutionary effects from effects of the initial conditions. We have collected a series of spectropolarimetric data of $\sim60$ HAeBe stars in three very young clusters (NGC 2244, NGC 2264 and NGC 6611), in which we discovered 3 magnetic stars \citep[][Alecian et al. in prep.]{alecian08b,alecian09a}. An analysis of the whole sample is in progress and will be published in the near future.

{ Finally three of the five magnetic stars constituing this sample belong to binary systems, therefore tidal braking could also be responsible of their slow rotation. However the three systems are not synchronised, and are wide ($a/R>10$, where $a$ is the semi-major axis of the orbit, and $r$ is the radius of the star). According to \citet{zahn77}, in the case of wide binaries, the evolution of the angular velocity tends towards an evolution in absence of tidal braking, meaning that tidal braking in wide binaries are negligible, especially during the first stages of stellar evolution. It implies that in such wide binaries, the synchronisation would occur on timescales of the order of the main-sequence lifetime, consistent with the observation of our three very young and not-synchronised systems. We therefore think that it is unlikely that tidal torques have been efficient enough to brake these young stars. Furthermore, in these systems, the magnetic component is systematically rotating slower than its non-magnetic companion. One of both hypotheses, involving magnetic field, that we exposed above, appear therefore more reasonable to explain the slow rotation of the magnetic Herbig Ae/Be stars.
}

%________________________________________________________________

\subsection{Angular momentum evolution of the normal sample}

First, we find that whatever the choice of the internal angular momentum treatment (solid body rotation or constant specific angular momentum), { the results are the same. The plots of the angular momentum as a function of age show a clear trend at all masses. We checked that below 5 \msun\ this trend can be fully explained with the age-mass correlation that our sample shows at lower mass. On the contrary, we find that above 5 \msun, the observed trend is real: the angular momentum of the stars decreases as the age increases.}

{ We have compared the rotation rates \wsini\ of our ZAMS samples with those of the normal MS intermediate-mass stars in the CPRV catalogue \citep{glebocki00}.} We find that in the case of a solid body rotation, whatever the mass, no significant differences are observed between ZAMS HAeBe and MS CPRV samples. { This result does not allow us to confirm the decrease of the angular momentum with age described above, but neither is it in contradiction with this observation (a KS- test can only reject the null hypothesis, it cannot confirm it).. Furthermore we note with the eye an excess of fast rotators in our ZAMS sample at high-mass compared to the MS sample.}

In the case of a constant specific angular momentum, when all masses are considered, it looks like too many fast HAeBe rotators are predicted. This trend seems to be observed only at high masses ($M>3.4$~\msun), which is confirmed with a KS test. The comparison of our ZAMS HAeBe sample with two other MS samples (RZG and $\alpha$ Per) confirms the trend at low-mass ($M<3.4$~\msun). Unfortunately, the RZG and $\alpha$Per have no or an insufficient number of high-mass stars ($M>3.4$\msun) to test if our conclusions are dependant on the choice of the MS sample.}

In conclusion, both analyses suggest that, whatever the internal angular momentum treatment, the lowest mass stars ($M<5$~M$_{\odot}$) evolve with constant angular momentum during the PMS phase, while at higher mass, the stars lose angular momentum. A larger sample of stars, especially at high-mass, is required to confirm this result.

\begin{figure}
\centering
\includegraphics[width=8.5cm]{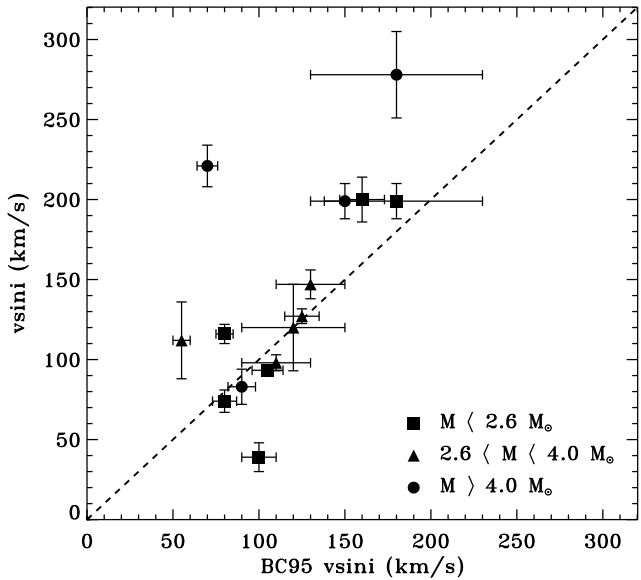}
\caption{\vsini\ of this work compared to the \vsini\ of BC95, for three mass ranges: $M<2.6$~M$_{\odot}$ (squares), $2.6<M<4$~M$_{\odot}$ (triangles), $M>4$~M$_{\odot}$ (dots). The dashed line is the first bisector.}
\label{fig:compv}
\end{figure}

\vspace{10pt}
A similar statistical analysis was performed in the past by \citet[][BC95 hereinafter]{bohm95} on a sample of 29 Herbig Ae/Be stars. Using MUSICOS (the ancestor of Narval and ESPaDOnS) echelle spectra with a resolution of $R = 38000$, they measured the \vsinis of all the stars using two methods. The first method is fitting many spectral lines with a convolution of a rotation profile and Gaussian of instrumental width. The second method is a comparison between many observed spectral lines and synthetic spectra of different rotational broadenings. They compared their \vsini\ to those of A and B stars in young open clusters.

They concluded that, if the star rotates as a solid body, low-mass HAeBe stars ($M<2.6$~M$_{\odot}$) lose angular momentum, intermediate mass HAeBe stars ($2.6<M<4$~M$_{\odot}$) evolve towards the ZAMS at constant angular momentum, while the high-mass HAeBe stars ($M>4$~M$_{\odot}$) gain angular momentum. Except for the intermediate-mass stars, these results are in disagreement with ours\footnote{We checked that, if we take the same mass limits as BC95, our conclusions remain unchanged.}.

By comparing our sample with that of BC95 we find 15 stars in common. In order to understand the discrepancy between our conclusions and the conclusions of BC95, we compared our \vsinis measurements to those of BC95. In Figure \ref{fig:compv} are plotted our \vsinis as a function of BC95's. We observe that the \vsinis measurements of the high mass stars from BC95 tend to be lower than ours, which could be the reason for the discrepancy. The reason of the differences obtained between BC95's \vsini\ and ours is not clear. It could come from the different treatment of the CS contamination. However independent \vsini\ determination from the same set of data than ours gives \vsini\ values consistent with ours \citep{folsom12}.

\citet{wolff04}, from their studies of PMS stars in Orion, concluded that intermediate-mass ($1.5 < M < 3.5$~M$_{\odot}$) stars lose angular momentum before they start the PMS radiative phase, then, during the PMS radiative evolution they evolve at constant angular momentum. In Fig. \ref{fig:hr} we observe that all the stars of our sample are experiencing the radiative phase of their PMS evolution. Therefore, our work confirms \citet{wolff04}'s conclusions.

\vspace{10pt}
Our work { suggests} that the more massive normal HAeBe stars ($M < 5$ M$_{\odot}$) are losing angular momentum as they evolve towards the ZAMS. These stars have no strong magnetic fields, and they are most likely not members of close binary systems. In order to understand this loss, and also the observed difference between high and low-mass HAeBe stars, we have searched for signs of winds in our spectra, especially in the Balmer lines, in the He~I~5875 \angs line, and in the multiplet 42 of Fe II. We also searched for signs of winds in UV spectra using the MAST archive, either in the C IV 1548/1550 \angs lines in massive stars or in the Mg II h \& k doublet in the HAe stars, that can display strong signs of winds. Of the 11 stars above 5 M$_{\odot}$, we find that 8 of them show either clear P Cygni profiles or a strong blueshifted absorption component in one or many of the searched lines, revealing the presence of a wind in the close stellar environment. Winds seem to be largely present among massive HAeBe stars and could be at the origin of their loss of angular momentum.

We have argued before that winds could also be partly responsible for the loss of angular momentum in magnetic stars. However, when coupled to strong magnetic fields, winds are expected to be much more efficient at carrying off angular momentum than in stars hosting faint or no magnetic fields. Therefore, even if all HAeBe stars have winds and are losing angular momentum, we still expect magnetic stars to rotate more slowly than non-magnetic stars, which
is consistent with our observations. .

In the case of low-mass HAeBe stars, the predicted ZAMS \vsinis values are similar to those observed in MS stars. However we should note that most of the sample stars at low mass have completed more than 50\% of the PMS phase. We therefore cannot conclude anything about the evolution of angular momentum during the first half of the PMS, for low-mass HAeBe stars. There are different ways to fill this gap: one way is to study HAeBe stars in very young clusters, such as NGC 6611, NGC 2244 and NGC 2264, that have ages from less than 1 Myr to $\sim2.5$ Myr. The results of our analysis of the spectropolarimetric data that we obtained for these clusters will be published in a future paper, and will discuss that point. Another way to fill this gap is to study intermediate-mass T Tauri stars, which have masses between 1.5 and $\sim3.5$ M$_{\odot}$ , and are evolving through the convective phase of their PMS evolution.  These stars have not been included into our analysis. An ESPaDOnS program is in progress in order to investigate these stars, and complete our knowledge of PMS angular momentum evolution at intermediate mass.

%
%______________________________________________________________

\section*{Acknowledgments}

We are very grateful to O. Kochukhov, who provided his BINMAG1 code. We thank Benjamin Montesinos, the referee, which led to major improvements in the paper. EA has been supported by the Marie Curie FP6 program, and the Centre National d'Etudes Spatiales. GAW and JDL acknowledge support from the Natural Science and Engineering Research Council of Canada (NSERC). GAW has also been supported by the DND Academic Research Programme (ARP). This research has made use of the SIMBAD database and the VizieR catalogue access tool, operated at CDS, Strasbourg (France), of the NASAs Astrophysics Data System, and of the Washington Double Star Catalog maintained at the U.S. Naval Observatory.

%
%______________________________________________________________

%\nocite{*}
\bibliographystyle{mn2e}
\bibliography{haebe-rot}

\begin{thebibliography}{}

\bibitem[\protect\citeauthoryear{{Abt} \& {Morrell}}{{Abt} \&
  {Morrell}}{1995}]{abt95}
{Abt} H.~A.,  {Morrell} N.~I.,  1995, \apjs, 99, 135

\bibitem[\protect\citeauthoryear{{Alecian}, {Catala}, {Wade}, {Bagnulo},
  {B{\"o}hm}, {Bouret}, {Donati}, {Folsom}, {Grunhut}, {Landstreet}, {Marsden},
  {Petit}, {Ramirez} \& {Silvester}}{{Alecian} et~al.}{2009a}]{alecian09a}
{Alecian} E.,  {Catala} C.,  {Wade} G.~A.,  {Bagnulo} S.,  {B{\"o}hm} T.,
  {Bouret} J.-C.,  {Donati} J.-F.,  {Folsom} C.,  {Grunhut} J.,  {Landstreet}
  J.~D.,  {Marsden} S.~C.,  {Petit} P.,  {Ramirez} J.,    {Silvester} J.,
  2009a, in {C.~Neiner \& J.-P.~Zahn} ed., EAS Publications Series Vol.~39 of
  EAS Publications Series, {Magnetism in Herbig Ae/Be stars and the link to the
  Ap/Bp stars}.
pp 121--132

\bibitem[\protect\citeauthoryear{{Alecian}, {Catala}, {Wade}, {Donati},
  {Petit}, {Landstreet}, {B{\"o}hm}, {Bouret}, {Bagnulo}, {Folsom}, {Grunhut}
  \& {Silvester}}{{Alecian} et~al.}{2008a}]{alecian08a}
{Alecian} E.,  {Catala} C.,  {Wade} G.~A.,  {Donati} J.,  {Petit} P.,
  {Landstreet} J.~D.,  {B{\"o}hm} T.,  {Bouret} J.,  {Bagnulo} S.,  {Folsom}
  C.,  {Grunhut} J.,    {Silvester} J.,  2008a, \mnras, 385, 391

\bibitem[\protect\citeauthoryear{{Alecian}, {Wade}, {Catala}, {Bagnulo},
  {B{\"o}hm}, {Bohlender}, {Bouret}, {Donati}, {Folsom}, {Grunhut} \&
  {Landstreet}}{{Alecian} et~al.}{2008b}]{alecian08b}
{Alecian} E.,  {Wade} G.~A.,  {Catala} C.,  {Bagnulo} S.,  {B{\"o}hm} T.,
  {Bohlender} D.,  {Bouret} J.,  {Donati} J.,  {Folsom} C.~P.,  {Grunhut} J.,
   {Landstreet} J.~D.,  2008b, \aap, 481, L99

\bibitem[\protect\citeauthoryear{{Alecian}, {Wade}, {Catala}, {Bagnulo},
  {B{\"o}hm}, {Bouret}, {Donati}, {Folsom}, {Grunhut} \&
  {Landstreet}}{{Alecian} et~al.}{2009b}]{alecian09b}
{Alecian} E.,  {Wade} G.~A.,  {Catala} C.,  {Bagnulo} S.,  {B{\"o}hm} T.,
  {Bouret} J.,  {Donati} J.,  {Folsom} C.~P.,  {Grunhut} J.,    {Landstreet}
  J.~D.,  2009b, \mnras, pp 1250--+

\bibitem[\protect\citeauthoryear{{Alecian}, {Wade}, {Catala}, {Grunhut},
  {B{\"o}hm}, {Bouret}, {Flood}, {Folsom}, , {Landstreet}, {Marsden} \&
  {Waite}}{{Alecian} et~al.}{2012}]{paperi}
{Alecian} E.,  {Wade} G.~A.,  {Catala} C.,  {Grunhut} J.~D.,  {B{\"o}hm} T.,
  {Bouret} J.,  {Flood} J.,  {Folsom} C.,   {Landstreet} J.~D.,  {Marsden}
  S.~C.,    {Waite} I.~A.,  2012, \mnras, in press

\bibitem[\protect\citeauthoryear{{Alonso-Albi}, {Fuente}, {Bachiller}, {Neri},
  {Planesas}, {Testi}, {Bern{\'e}} \& {Joblin}}{{Alonso-Albi}
  et~al.}{2009}]{alonso09}
{Alonso-Albi} T.,  {Fuente} A.,  {Bachiller} R.,  {Neri} R.,  {Planesas} P.,
  {Testi} L.,  {Bern{\'e}} O.,    {Joblin} C.,  2009, \aap, 497, 117

\bibitem[\protect\citeauthoryear{{Auri{\`e}re}, {Wade}, {Silvester},
  {Ligni{\`e}res}, {Bagnulo} \& {Bale}}{{Auri{\`e}re} et~al.}{2007}]{auriere07}
{Auri{\`e}re} M.,  {Wade} G.~A.,  {Silvester} J.,  {Ligni{\`e}res} F.,
  {Bagnulo} S.,    {Bale} K. e.~a.,  2007, \aap, 475, 1053

\bibitem[\protect\citeauthoryear{{Behrend} \& {Maeder}}{{Behrend} \&
  {Maeder}}{2001}]{behrend01}
{Behrend} R.,  {Maeder} A.,  2001, \aap, 373, 190

\bibitem[\protect\citeauthoryear{{Belikov}}{{Belikov}}{1995}]{belikov95}
{Belikov} A.~N.,  1995, Bulletin d'Information du Centre de Donnees Stellaires,
  47, 9

\bibitem[\protect\citeauthoryear{{B{\"o}hm} \& {Catala}}{{B{\"o}hm} \&
  {Catala}}{1995}]{bohm95}
{B{\"o}hm} T.,  {Catala} C.,  1995, \aap, 301, 155

\bibitem[\protect\citeauthoryear{{Calvet}, {Muzerolle}, {Brice{\~n}o},
  {Hern{\'a}ndez}, {Hartmann}, {Saucedo} \& {Gordon}}{{Calvet}
  et~al.}{2004}]{calvet04}
{Calvet} N.,  {Muzerolle} J.,  {Brice{\~n}o} C.,  {Hern{\'a}ndez} J.,
  {Hartmann} L.,  {Saucedo} J.~L.,    {Gordon} K.~D.,  2004, \aj, 128, 1294

\bibitem[\protect\citeauthoryear{{Donati}, {Semel}, {Carter}, {Rees} \&
  {Collier Cameron}}{{Donati} et~al.}{1997}]{donati97}
{Donati} J.-F.,  {Semel} M.,  {Carter} B.~D.,  {Rees} D.~E.,    {Collier
  Cameron} A.,  1997, MNRAS, 291, 658

\bibitem[\protect\citeauthoryear{{Dullemond}, {Dominik} \& {Natta}}{{Dullemond}
  et~al.}{2001}]{dullemond01}
{Dullemond} C.~P.,  {Dominik} C.,    {Natta} A.,  2001, \apj, 560, 957

\bibitem[\protect\citeauthoryear{{Finkenzeller} \& {Mundt}}{{Finkenzeller} \&
  {Mundt}}{1984}]{finkenzeller84}
{Finkenzeller} U.,  {Mundt} R.,  1984, \aaps, 55, 109

\bibitem[\protect\citeauthoryear{{Folsom}, {Bagnulo}, {Wade}, {Alecian},
  {Landstreet}, {Marsden} \& {Waite}}{{Folsom} et~al.}{2012}]{folsom12}
{Folsom} C.~P.,  {Bagnulo} S.,  {Wade} G.~A.,  {Alecian} E.,  {Landstreet}
  J.~D.,  {Marsden} S.~C.,    {Waite} I.~A.,  2012, ArXiv e-prints

\bibitem[\protect\citeauthoryear{{Folsom}, {Wade}, {Kochukhov}, {Alecian},
  {Catala}, {Bagnulo}, {B{\"o}hm}, {Bouret}, {Donati}, {Grunhut}, {Hanes} \&
  {Landstreet}}{{Folsom} et~al.}{2008}]{folsom08}
{Folsom} C.~P.,  {Wade} G.~A.,  {Kochukhov} O.,  {Alecian} E.,  {Catala} C.,
  {Bagnulo} S.,  {B{\"o}hm} T.,  {Bouret} J.,  {Donati} J.,  {Grunhut} J.,
  {Hanes} D.~A.,    {Landstreet} J.~D.,  2008, \mnras, 391, 901

\bibitem[\protect\citeauthoryear{{Glebocki} \& {Stawikowski}}{{Glebocki} \&
  {Stawikowski}}{2000}]{glebocki00}
{Glebocki} R.,  {Stawikowski} A.,  2000, Acta Astronomica, 50, 509

\bibitem[\protect\citeauthoryear{{Herbig}}{{Herbig}}{1960}]{herbig60}
{Herbig} G.~H.,  1960, \apjs, 4, 337

\bibitem[\protect\citeauthoryear{{Ligni\`eres}, {Catala} \&
  {Mangeney}}{{Ligni\`eres} et~al.}{1996}]{lignieres96}
{Ligni\`eres} F.,  {Catala} C.,    {Mangeney} A.,  1996, \aap, 314, 465

\bibitem[\protect\citeauthoryear{{Mason}, {Breeveld}, {Much}, {Carter},
  {Cordova}, {Cropper}, {Fordham}, {Huckle}, {Ho}, {Kawakami}, {Kennea},
  {Kennedy}, {Mittaz}, {Pandel}, {Priedhorsky}, {Sasseen}, {Shirey}, {Smith} \&
  {Vreux}}{{Mason} et~al.}{2001}]{mason01}
{Mason} K.~O.,  {Breeveld} A.,  {Much} R.,  {Carter} M.,  {Cordova} F.~A.,
  {Cropper} M.~S.,  {Fordham} J.,  {Huckle} H.,  {Ho} C.,  {Kawakami} H.,
  {Kennea} J.,  {Kennedy} T.,  {Mittaz} J.,  {Pandel} D.,  {Priedhorsky} W.~C.,
   {Sasseen} T.,  {Shirey} R.,  {Smith} P.,    {Vreux} J.,  2001, \aap, 365,
  L36

\bibitem[\protect\citeauthoryear{{Meeus}, {Waters}, {Bouwman}, {van den
  Ancker}, {Waelkens} \& {Malfait}}{{Meeus} et~al.}{2001}]{meeus01}
{Meeus} G.,  {Waters} L.~B.~F.~M.,  {Bouwman} J.,  {van den Ancker} M.~E.,
  {Waelkens} C.,    {Malfait} K.,  2001, \aap, 365, 476

\bibitem[\protect\citeauthoryear{{Mer{\'{\i}}n}, {Montesinos}, {Eiroa},
  {Solano} \& {Mora}}{{Mer{\'{\i}}n} et~al.}{2004}]{merin04}
{Mer{\'{\i}}n} B.,  {Montesinos} B.,  {Eiroa} C.,  {Solano} E.,    {Mora} A.,
  2004, \aap, 419, 301

\bibitem[\protect\citeauthoryear{{Michaud}}{{Michaud}}{1970}]{michaud70}
{Michaud} G.,  1970, \apj, 160, 641

\bibitem[\protect\citeauthoryear{{Morel}}{{Morel}}{1997}]{morel97}
{Morel} P.,  1997, A\&AS, 124, 597

\bibitem[\protect\citeauthoryear{{Palla} \& {Stahler}}{{Palla} \&
  {Stahler}}{1993}]{palla93}
{Palla} F.,  {Stahler} S.~W.,  1993, ApJ, 418, 414

\bibitem[\protect\citeauthoryear{{Press}, {Teukolsky}, {Vetterling} \&
  {Flannery}}{{Press} et~al.}{1992}]{press92}
{Press} W.~H.,  {Teukolsky} S.~A.,  {Vetterling} W.~T.,    {Flannery} B.~P.,
  1992, {Numerical recipes in C. The art of scientific computing}

\bibitem[\protect\citeauthoryear{{Prosser}}{{Prosser}}{1992}]{prosser92}
{Prosser} C.~F.,  1992, \aj, 103, 488

\bibitem[\protect\citeauthoryear{{Royer}, {Zorec} \& {G{\'o}mez}}{{Royer}
  et~al.}{2007}]{royer07}
{Royer} F.,  {Zorec} J.,    {G{\'o}mez} A.~E.,  2007, \aap, 463, 671

\bibitem[\protect\citeauthoryear{{St{\c e}pie{\'n}}}{{St{\c
  e}pie{\'n}}}{2000}]{stepien00}
{St{\c e}pie{\'n}} K.,  2000, \aap, 353, 227

\bibitem[\protect\citeauthoryear{{Strom}, {Strom}, {Yost}, {Carrasco} \&
  {Grasdalen}}{{Strom} et~al.}{1972}]{strom72}
{Strom} S.~E.,  {Strom} K.~M.,  {Yost} J.,  {Carrasco} L.,    {Grasdalen} G.,
  1972, \apj, 173, 353

\bibitem[\protect\citeauthoryear{{Svechnikov} \& {Bessonova}}{{Svechnikov} \&
  {Bessonova}}{1984}]{svechnikov84}
{Svechnikov} M.~A.,  {Bessonova} L.~A.,  1984, Bulletin d'Information du Centre
  de Donnees Stellaires, 26, 99

\bibitem[\protect\citeauthoryear{{Th{\'e}}, {de Winter} \& {Perez}}{{Th{\'e}}
  et~al.}{1994}]{the94}
{Th{\'e}} P.~S.,  {de Winter} D.,    {Perez} M.~R.,  1994, \aaps, 104, 315

\bibitem[\protect\citeauthoryear{{Vieira}, {Corradi}, {Alencar}, {Mendes},
  {Torres}, {Quast}, {Guimar{\~a}es} \& {da Silva}}{{Vieira}
  et~al.}{2003}]{vieira03}
{Vieira} S.~L.~A.,  {Corradi} W.~J.~B.,  {Alencar} S.~H.~P.,  {Mendes}
  L.~T.~S.,  {Torres} C.~A.~O.,  {Quast} G.~R.,  {Guimar{\~a}es} M.~M.,    {da
  Silva} L.,  2003, \aj, 126, 2971

\bibitem[\protect\citeauthoryear{{Wolff}, {Strom} \& {Hillenbrand}}{{Wolff}
  et~al.}{2004}]{wolff04}
{Wolff} S.~C.,  {Strom} S.~E.,    {Hillenbrand} L.~A.,  2004, \apj, 601, 979

\bibitem[\protect\citeauthoryear{{Zahn}}{{Zahn}}{1977}]{zahn77}
{Zahn} J.-P.,  1977, \aap, 57, 383

\end{thebibliography}

%
%______________________________________________________________

%
%______________________________________________________________

\appendix
\section{Spectral type - mass relation}

\begin{figure}
\centering
\includegraphics[width=7.5cm]{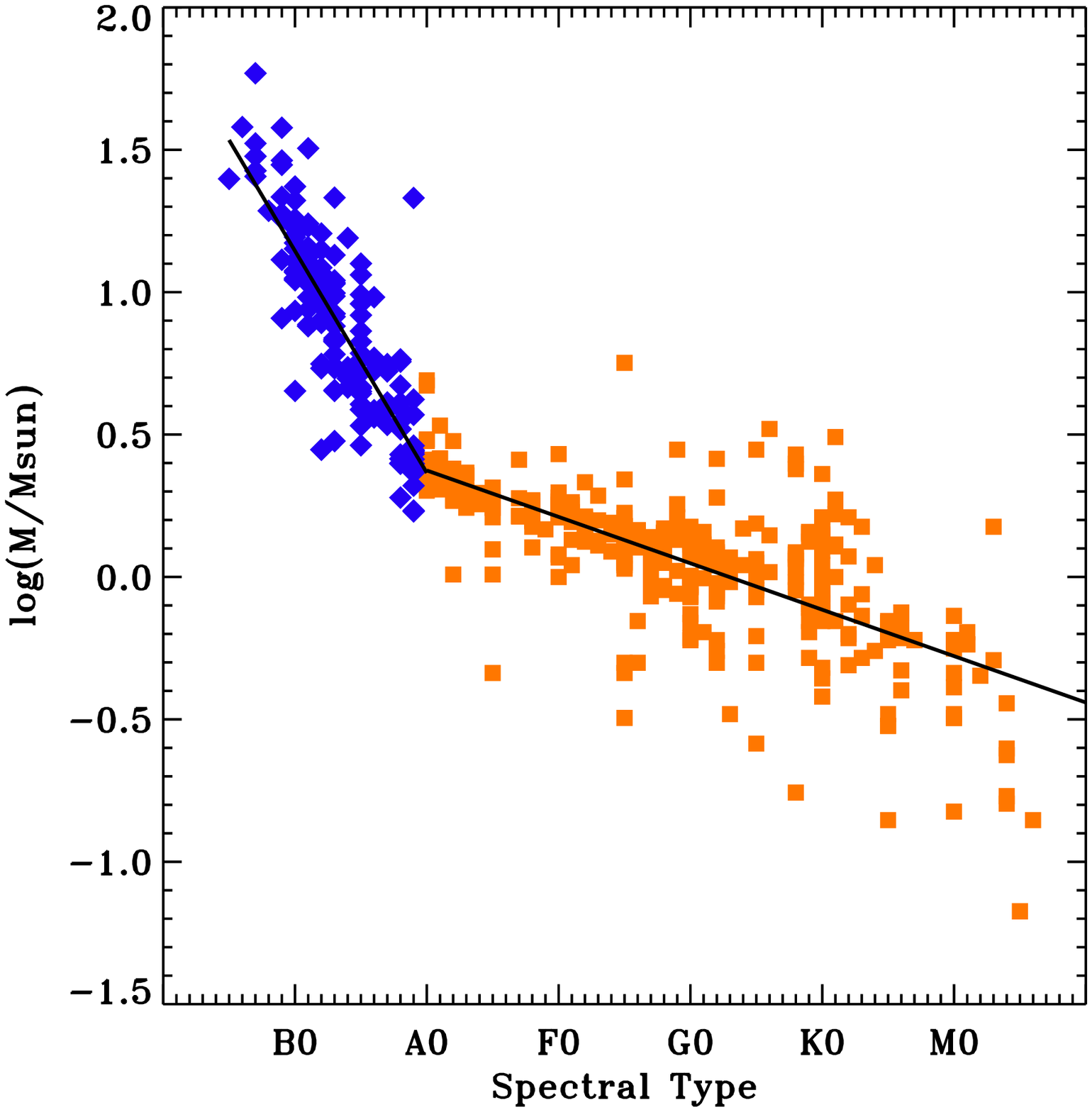}
\caption{Masses as a function of spectral types for the stars of the SMaC catalogue \citep{belikov95}. Both linear fits for spectral types earlier and later than A0 are overplotted.}
\label{fig:mcalib}
\end{figure}

In order to determine the masses of main sequence stars, we used the Stellar Mass Catalogue (SMaC) of \citet{belikov95} that contains the dynamical determination of masses of binary systems of luminosity class V, as well as their spectral types. In Fig. \ref{fig:mcalib} are plotted the masses of the stars of this catalogue as a function of their spectral type. We observe that two linear trends can be drawn below and above spectral type A0. We performed a linear fit to both trends and obtained the following formulae:
\begin {itemize}
\item for spectral types earlier than A0
\begin{equation}
\;\;\;\;\;\;\log\left(\frac{M}{M_{\odot}}\right) = 1.9 - 0.078\times spt
\end{equation}
\item for spectral types later than A0
\begin{equation}
\;\;\;\;\;\;\log\left(\frac{M}{M_{\odot}}\right) = 0.70 - 0.016\times spt
\end{equation}
\end{itemize}
where $spt$ increases by 1 for every spectral subtype, beginning at 0 for spectral type O0.  Thus for spectral type B0 spt is 10, for A0 it is 20, and so on.

\section{Mass - radius relation}

\begin{figure}
\centering
\includegraphics[width=7.5cm]{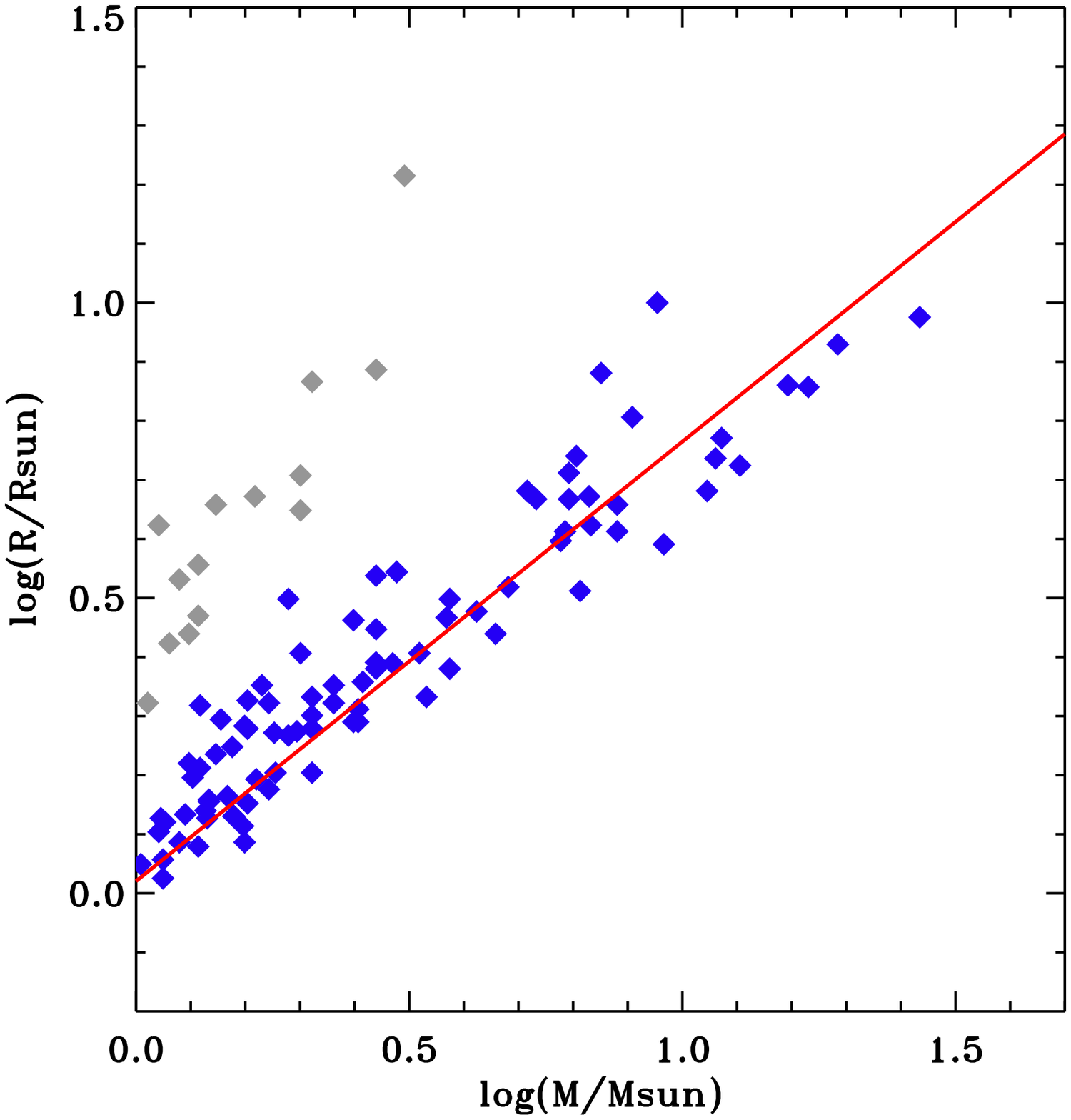}
\caption{Radii as a function of masses for the stars of the catalogue of \citep{svechnikov84}. The linear fit is overplotted. The rejected data points are plotted in light grey.}
\label{fig:rcalib}
\end{figure}

In order to determine the radius of main sequence stars, we used the Catalogue of orbital elements, mass and luminosities of close double stars of \citet{svechnikov84} that contains the mass and radius determinations of both components of eclipsing binary systems of luminosity class V. In Fig. \ref{fig:rcalib} are plotted the radii of the stars of this catalogue as a function of their masses for masses larger than 1~\msun. We observe a linear trend at the lower limit of the data set, and a scattering that increases towards low mass. We suggest that the data points that are situated well above the general trend, have a wrong published luminosity class. These points are considered as outsiders, and have not been included into the fit. We performed a linear fit to the remaining data points and obtained the following formula:
\begin{equation}
\;\;\;\;\;\;\log\left(\frac{R}{R_{\odot}}\right) = 0.021 + 0.074*\log\left(\frac{M}{M_{\odot}}\right)
\end{equation}

%
%______________________________________________________________

\end{document}